\documentclass{ws-p9-75x6-50}
\def\msun{{M_\odot}}
\newcommand{\mpc}{{\hbox {$h^{-1}$}{\rm Mpc}} }
\def\kms{\,{\rm {km\, s^{-1}}}}
\def\be{\begin{equation}}
\def\ee{\end{equation}}
\def\bea{\begin{eqnarray}}
\def\eea{\end{eqnarray}}

\begin{document}

\title{Warm Dark Matter Model of Galaxy Formation}

\author{Y.P. Jing}

\address{Shanghai Astronomical Observatory, Partner Group of MPI f\"ur Astrophysik, Nandan RD 80, Shanghai 200030, China \\E-mail: ypjing@center.shao.ac.cn}


\maketitle\abstracts{ Cold Dark Matter (CDM) models of
galaxy formation had been remarkably successful to explain a number of
observations in the past decade. However, with both the theoretical
modeling and the observations being improved, CDM models have been
very recently shown to have excessive clustering on the sub-galactic
scale.  Here I discuss a solution, based on our high-resolution
numerical simulations, to this outstanding problem by considering Warm
Dark Matter (WDM). Our results show that the over-clustering problem
on sub-galactic scales can be overcome by WDM models, and all the
advantages of CDM models are preserved by WDM models. Therefore, the
WDM model will become an interesting alternative to the well-studied
CDM models}

\section{Introduction}

Cold Dark Matter (CDM) models have been shown very successful to
explain many observations of galaxies on scales of about one $\mpc$ to
a few hundred $\mpc$. But such models probably predict over-clustering
on smaller scales, as recent high-resolution simulations (Moore et
al. 1999, Klypin et al. 1999, Jing \& Suto 2000) have shown. The halo
density profiles in these simulations are steeper than those inferred
from the rotation curves of low surface brightness (LSB) galaxies, and there
are too many sub-halos within galactic halos when compared to the
observed number of satellite galaxies around the Milky Way. There is
also additional evidence for such overclustering from, e.g.  the
luminosity function of dwarf galaxies. Although some of these
discrepancies may be resolved by introducing additional astrophysical
processes (Bullock et al. 2000) and some others by properly
interpreting the observations (van den Bosch \& Swaters 2000), there
are attempts to resolve the discrepancies by revisiting the assumption
about the dark matter (DM). A list of the candidates for replacing CDM
proposed since the summer of 1999 includes self-interacting DM, warm
dark matter (WDM), repulsive DM, fuzzy DM, annihilating DM etc (see
Dav\`e et al. 2000 for references).  In this talk, I will present an
extensive study for a warm dark matter model using high resolution
N-body simulations. Our results will show that the WDM model is in
good agreement with the observational data without resorting to
not-well-understood astrophysical processes. A complete description of
the study appeared in our recent paper submitted to the Astrophyical
Journal (Jing 2000).

\section{Model and Simulations}
We consider a model dominated by WDM with the matter density
$\Omega_0=0.3$, the cosmological constant $\lambda_0=0.7$, and the
Hubble constant $H_0= 100h = 67 {\rm km s^{-1} Mpc^{-1}}$. The
primordial power spectrum is $\propto k$, and the transfer function is
taken from Bardeen et al. (1986) with a zero baryon content. The
free-streaming cutoff parameter $R_f=0.1 \mpc$ is adopted, which is
also consistent with the Ly-$\alpha$ forest observations
(Narayanan et al. 2000).  The linear power spectrum is normalized
so that the current {\rm rms} linear density perturbation within a sphere
of radius $8 \mpc$ is 1. Because all the parameters except $R_f$ have usually
been assumed for the low-density flat CDM (LCDM) model which best fits
observations on scales of $1\mpc$ and up, this WDM is expected to fit
these observations as well since the free-steaming of the
warm dark matter has little effect on these scales. Thus our study will
focus on the properties on galactic and sub-galactic scales, where the
free-streaming effect of the WDM is expected to become significant. To
single out the free-streaming effect, we compare the results of the WDM
to those of the LCDM, which in most cases can effectively eliminate
the numerical artifacts.

We have run a large set of cosmological simulations with box sizes of
$12.5\mpc$, $25\mpc$ and $50\mpc$ respectively. For each box size,
three realizations are produced and $128^3$ particles are adopted for
each model. The same initial phases are used for the WDM model and for
the LCDM model. We have selected five WDM halos of $\sim 1000$
particles for each boxsize as well as the corresponding halos
in the LCDM simulations. The virial mass of these halos is $7\times
10^{10}\msun$, $6\times 10^{11}\msun$, and $5\times 10^{12}\msun$
respectively from the small to the large boxsizes.
We then use the Nested-Grid-${\rm P^3M}$ (Jing \& Suto 2000) to
simulate these halos with a much higher resolution. A total of $\sim 7\times 10^5$ particles are used for simulating each halo, with $\sim 5\times
10^5$ particles (small) for the high-resolution region and $\sim 2\times
10^5$ (massive) for the coarse-resolution region. About $3\times 10^5$
particles from the high-resolution region will end up in the virialized region
of the halo. Using the coarse (massive) particles can properly account
for the tidal force which is important for forming the
internal structures of the halos. A detailed account about the
simulation technique can be found in Jing \& Suto (2000).

\section{Results}
We have made a very detailed analysis both for the cosmological
simulations and for the high-resolution halo simulations. These
results were presented in the talk, but can not be accommodated in
this proceedings paper because of the limited space. A detailed
account of these results can be found in our journal paper (Jing
2000). Here we just highlight a few interesting results.

\begin{figure}
\psfig{figure=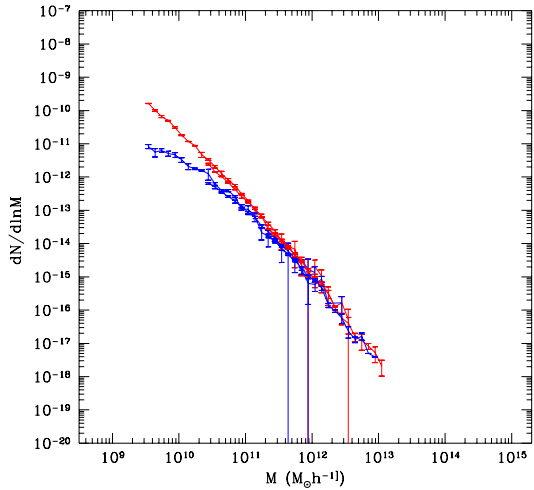,height=3in}
\caption{The differential mass function of dark matter halos at redshift $z=4$ in the WDM model (lower) and in the LCDM model (upper). \label{fig:fig1}}
\end{figure}

In Figure 1 we show the differential mass function of dark matter
halos in the WDM model as well as in the LCDM model at redshift
$z=4$. The mass function is defined as the mean number density of
halos within a unit logarithmic interval of halo mass. Because of the
free-streaming motion of the warm dark matter, the halo abundance in
the WDM model is smaller than in the LCDM model at sub-galactic
scales. The WDM halos are about 2 times and 4 times less abundant at
$M= 10^{11}\msun$ and $M=2\times 10^{10}\msun$ respectively. A crucial
test for this effect would be the hydrogen content of the damped
Ly-$\alpha$ systems. Since the halo density is reduced only by a
factor of two on the relevant scales, the WDM is well consistent with
the observations of the damped Ly-$\alpha$ systems (see Mo \&
Miralda-Escude 1994, Ma et al. 1997)

The density profiles of dark matter halos are presented in Figure
2. We found that the profiles of the halos less massive than $5\times
10^{11}\msun$ are significantly flatter in the WDM model. Especially
the halos with the mass similar to the those of the dwarf galaxies
have formed cores at their centers, which could be potentially
important for explaining the slowly rising rotation curves observed
for the LSB galaxies. Our comparison of the halo circular velocity
with the observed rotation curves shows that the LCDM model could be
ruled out for its too steep density profiles, and the WDM model is
consistent with the observations of LSB galaxies.

Now let's consider sub-halos within virialized halos. The smoothed
density of each dark matter particle is estimated in the way used in
Smoothed Particle Hydrodynamics simulations. The average of the
smoothed density is calculated for each radial shell, and the
particles with the smoothed density five times above the radially averaged
value are identified. These identified particles are grouped with the
friend-of-friend algorithm of a bonding length equal to one tenth of
the global mean particle separation. We find that the resulted
catalogue of the sub-halos is quite robust against the parameters we
have taken. Figure 3 shows the number of sub-halos with the circular
velocity larger than $v_c$. There are much fewer sub-halos in the WDM
halos than in the LCDM halos.  To compare with obsrevations, we plot
the observed abundance of the satellite galaxies within the Milky Way
which has a circular velocity of $220\kms$. The number of sub-halos
within a Milky Way like halo, which should be in between the curves of
$v_{c,host}=270\kms$ and of $v_{c,host}=135\kms$, can be
readily read out for the two dark matter models.  The LCDM halos have
too many halos as many previous studies have pointed out, but the WDM
halos have a comfortable amount of sub-halos which is in good
agreement with the observations.  We noted that the our results for
the LCDM are in good agreement, for the mass range plotted in the
figure, with the result of Moore et al.(1999).

\begin{figure*}
\psfig{figure=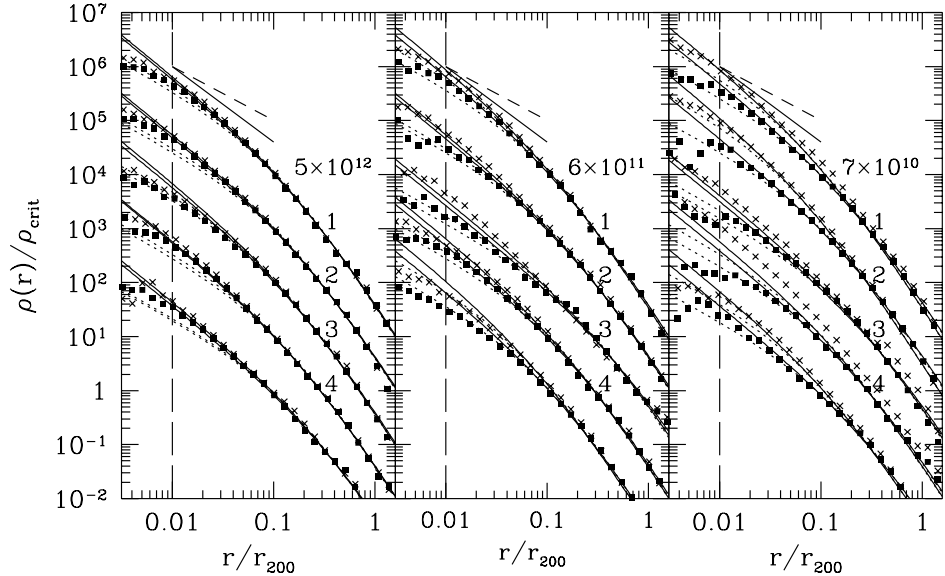,height=3in}
\caption{The density profiles of the dark matter halos in the WDM
model (squares) and in the LCDM model (crosses). \label{fig:fig2}}
\end{figure*}

\begin{figure*}
\psfig{figure=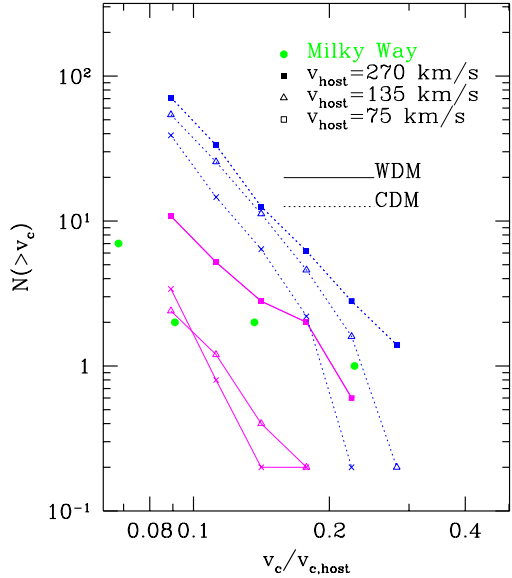,height=3in}
\caption{The number of sub-halos with the circular velocity larger
than $v_c$ as a function of $v_c/v_{c,host}$ where $v_{c,host}$ is the
circular velocity of the virialized host halo where the sub-halos
reside.  \label{fig:fig3}}
\end{figure*}

In summary, we have presented a first, very detailed simulation study
for the WDM model. Our results show that the model predicts enough
halos at high redshift $z\approx 4$ to be consistent with the
observations of damped Ly-$\alpha$ systems. This model is also
consistent with the clustering of the Ly-$\alpha$ absorption lines
(Narayanan et al. 2000). The density profiles of the halos are
significantly flatter than in the LCDM model for halo mass less than
$\sim 10^{11}\msun$, bringing about good agreement with the recent
high-resolution observation of the rotation curves of LSB galaxies by
Swaters et al.(2000). In contrast, we found that the CDM halos are NOT
consistent with the observation of Swaters et al.. There are
significantly fewer subhalos in WDM halos than in LCDM halos, and the
number of subhalos within Milky Way like halos in the WDM model agrees
very well with the observed number of satellite galaxies around the
Milky Way without resorting to poorly-understood astrophysical
processes. Since there are much fewer sub-halos, we expect that the
over-cooling problem on the galactic scales can be alleviated and
large galactic disks can be formed, which has been a serious
difficulty for CDM models. All these attractive features of the WDM
model warrant further detailed studies of this model.

The work is supported by the One-Hundred-Talent Program and by 
NKBRSF-G19990754.

\end{document}